\documentclass[preprint,12pt]{elsarticle}

\usepackage{amsmath}
\usepackage{amsfonts}
\usepackage{amssymb}
\usepackage{graphicx}

\usepackage{setspace}
\usepackage{hyperref}

\usepackage[latin1]{inputenc}

\newcommand{\dd}{\mathrm{d}}

\usepackage{amssymb}

\journal{Journal of Non-Newtonian Fluid Mechanics}

\graphicspath{{figs/}}

\begin{document}

\begin{frontmatter}

\title{{\bf Thermal Chaotic Mixing of Power-Law Fluids in a Mixer with Alternately-Rotating Walls }}

\author[ky]{Kamal El Omari\corref{cor1}}
\ead{kamal.elomari@univ-pau.fr}
\author[ky]{Yves Le Guer}
\ead{yves.leguer@univ-pau.fr}

\address[ky]{{Laboratoire de Thermique Energétique et Procédés (LaTEP)} \\
	{ Université de Pau et des Pays de l'Adour (UPPA)}	\\
	{IFR, rue Jules Ferry, BP. 7511, 64075 Pau cedex - France}}

\cortext[cor1]{Corresponding author, Tel: +33 5 59 40 71 47, Fax: +33 5 59407160}

\begin{abstract}

 We investigate the enhancement of both mixing and heat transfer in a two-rod mixer for highly-viscous non-Newtonian fluids. The mixer is composed of two vertical circular rods in a cylindrical tank. Chaotic flows are obtained by imposing the temporal modulations of the rotational velocities of the walls. We study the effects of different stirring protocols, which lead to non-chaotic and chaotic flows, on the efficiency of both mixing and heat transfer for three different rheological fluid behaviors: shear-thinning, Newtonian and shear-thickening. For this purpose, we use statistical indicators that characterize the mean value of the fluid temperature and its homogenization. We find that chaotic mixing is suitable for shear-thickening fluids for which we observe a clear enhancement of the thermal mixing (heat extraction and homogenization). This is due to the increase in the apparent fluid viscosity in the vicinity of the rotating walls. This aspect confirms the relevance of chaotic mixing for highly-viscous fluids. 
\end{abstract}


\end{frontmatter}

\textbf{Nomenclature}
\begin{tabbing}
 \hspace{1.5cm}\=\kill
 $c_p$      	\> heat capacity ($J.kg^{-1}.K^{-1}$)\\
 $\overset{=}D$      	\> Rate of deformation tensor\\
 $k$      	\> consistency index ($Pa\ s^{n}$)\\ 
 $n$      	\> flow behavior index (dimensionless)\\ 
 $\vec{n}$      	\> normal-oriented unit vector\\
 $p$      	\> pressure ($Pa$) \\
 $R_3$      	\> tank radius ($m$)  \\
 $R_1, R_2$      	\> rod radii ($m$) \\
 $t$      	\> time ($s$)\\
 $T$      	\> temperature ($K$)\\
 $U$      	\> tangential velocity\\
 $\vec U$      	\> velocity field ($m.s^{-1}$)\\
 $X$      	\> rescaled dimensionless temperature\\
 \textbf{Dimensionless numbers}\\
 $Pe$      	\> Péclet number\\
 $Re$      	\> Reynolds number\\
 $Re_{PL}$      \> Generalized Reynolds number\\
 $T^*$      	\> dimensionless temperature\\
 \textbf{Greek symbols}\\
 $\varepsilon$      	\> eccentricity ($m$)\\
 $\phi$      	\> generic scalar variable\\
 $\Gamma$      	\> generic diffusion coefficient\\
 $\rho$      	\> fluid density ($kg.m^{-3}$)\\
 $\sigma$      	\> standard deviation\\
 $\tau$      	\> period of modulation ($s$)\\
 $\overset{=}{\tau}$      	\> viscous stress tensor\\
 $\Omega$      	\> angular velocity ($rd.s^{-1}$)\\
 \textbf{Subscript}\\
 $c$      	\> cell\\
 $m$      	\> mean\\
 $f$      	\> face of a cell\\
 \textbf{Superscript}\\
 $*$      	\> dimensionless\\
\end{tabbing}

\section*{Introduction}

Mixing processes are currently encountered in many practical engineering domains where enhancement of the heat, mass and momentum transfer are required. In the present study, mixing is achieved through the presence of a laminar chaotic flow that ensures the efficient stretching and folding of material lines (for a 2D flow in our case). The need for chaotic mixing is particularly interesting when high viscosity and/or shear-sensitive fluids are concerned \cite{ottino1989}. In this case, classical laminar 2D time-independent flows (i.e. regular flows)  are unable to give a good mixing performance and as a consequence, heat transport from the wall will be ineffective \cite{aref2002}. For example, such a situation exists for non-Newtonian molten polymers or polymer blends \cite{jana2004}. Thus, the processing of highly-viscous polymer-molecule networks can degrade in high shear regions of the flow. Also, undesired or ill-defined structures may be obtained for polymer blends. Furthermore, the problems of both heat dissipation and energy costs may become very important for highly-viscous fluids when classical impellers are used \cite{tanguy1997}.

Many research studies have concerned the usage of chaotic flows to promote heat transfer.  Mainly two classes of flow geometries have been encountered in the production of chaotic mixing at industrial scale: those that use rotating elements as eccentric cylinders \cite{mota2007,ganesan1997} and those that use multiple pipe bends \cite{peerhossaini1993,kumar2007b,Fellouah2010}. Since most of the fluids concerned in the industrial processes are non-Newtonian (ex: food and chemical industries), a growing interest is focused on the application and the study of chaotic advection to the mixing and heat transfer in these fluids. When dealing with non-Newtonian fluids, their rheological characteristics may influence the mixing efficiency. Using a computational approach, Niederkorn and Ottino \cite{niderkorn1994} showed that the shear-thinning character of the fluid decreases the rate and the extent of the mixing in a 2D, time-periodic flow between two eccentric cylinders. Anderson \textit{et al.} reported similar conclusions in 2D and 3D periodically agitated cavities \cite{anderson2000}. Baloch and Webster \cite{baloch2003} have investigated complex rotational 2D flows of viscous non-Newtonian fluids in a rotating vessel equipped with a single eccentrically positioned stirrer. They have shown that shear-thinning introduces larger shear-rates, but reduces localized rate-of-work maxima and global power. An extension of this work for 3D numerical flows and also with the consideration of two stirrers is presented in Sujatha and Webster \cite{Sujatha2003}.
 Connelly and Kokini simulated by a finite element method the flow in a screw mixer of four different fluids: a Newtonian, a shear-thinning, an elastic and a shear-thinning viscoelastic fluid. They analyzed the effects of these rheological behaviors separated and combined. They observed an increase of the size of dead zones due to the shear-thinning behavior \cite{connelly2004}. Kumar and Homsy used the Kaper and Wiggins geometric theory to investigate the effect of elasticity on the chaotic advection efficiency in a quasi-steady flow. They found that elasticity increases the rate of mixing by extending the zone concerned by chaotic flow \cite{Kumar1996}.
As for mixing of tracers, the heat transfer is strongly influenced by the local flow field in rheologically-complex fluids \cite{escudier2002}, thus it is important to study the impact of rheology on chaotic mixing and therefore on the heat transfer \cite{lester2009}.

The approach that this work takes is to analyze the combined effects of mixing and heat transfer in the case of shear-thinning and shear-thickening fluids and to compare the obtained results to those of Newtonian fluids. The flow in this mixer was previously studied for Newtonian fluids \cite{elomari2009a, elomari2009b}. It is characterized by unsteady velocity fields and streamlines that are compared for each of the three rheological behaviors. Different mixing indicators and statistical tools are also used to compare the heat transfer efficiency for the different complex fluids.

\section{Problem formulation}
\label{problemformulation}

\subsection{Geometry of the two-rod rotating mixer}
\begin{figure}[htbp]
 \centering
 \begin{center}
 \includegraphics[width=0.5\textwidth]{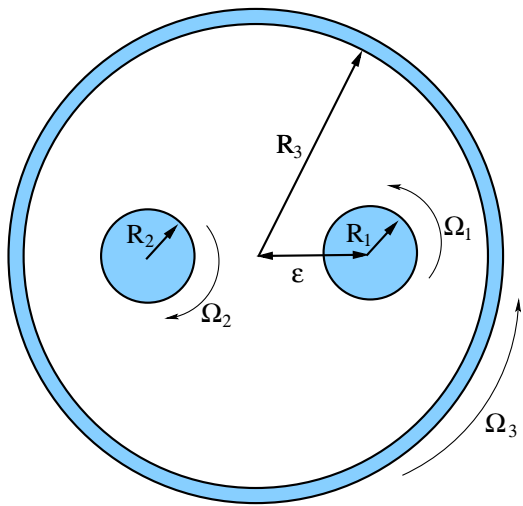}
\end{center}
 \caption{Sketch of the two-rod mixer.}
 \label{fig:mixer}
\end{figure}

A sketch of the mixer that was used in the study is presented in Fig. \ref{fig:mixer}. It is composed of two circular rods with equal radii, which are maintained vertical inside of a cylindrical tank (a bounded domain). The tank and the rods are heated and can rotate around their respective revolution axis. This geometry is similar to the two-roll mills that were studied in the literature by Price \textit{et al.} \cite{price2003}~; however, in our case, the outer cylinder has the ability to rotate. The rods and the tank rotate with an alternating or a continuous velocity modulation. They can also have different rotation directions. 
We have shown that for a Newtonian fluid \cite{elomari2009a} this two-rod mixer is suitable in order to obtain fully-chaotic flow without KAM regions, which is particularly interesting for industrial applications.

The geometry of the flow domain is characterized by the radii of the rod and the cylindrical tank, which are, respectively, $R_1 = R_2 = 10\ mm$ and $R_3 = 50\ mm$. The eccentricity of the rods is set to the value of $\varepsilon= 25\ mm$.

\subsection{Flow parameters: stirring protocols}
\label{strirringprot}
Chaotic mixing flows are produced by varying, with time, the angular velocity of the rods and the tank by using a sine-squared waveform. The stirring protocols that were studied are defined by two parameters: the respective rotation direction between the rods and the cylindrical tank, and the duration of the time periodic modulation. Among the three possible stirring configurations that correspond to different flow topologies, in this study, we chose a stirring configuration with co-rotating tank and rods. We have shown in previous work \cite{elomari2009b} that this configuration gives globally-chaotic flows without the existence of elliptic KAM regions in the Newtonian fluid. In the same study, two other types of stirring configurations were studied but it was demonstrated that they were less efficient.

For chaotic flows, two types of temporal modulation are considered for the stirring protocols: continuous (sine-squared modulation of the wall velocities) and non-continuous or alternating. In the case of the non-continuous modulation, the rods are stopped together for half a period while the outer tank is rotating and then, for the next half-period, the contrary occurs. These stirring protocols are illustrated in Fig. \ref{fig:modulation}. The maximum angular velocity is the same for the two types of modulation. The rods and the cylindrical tank follow the sine-squared modulations, which are defined by equations \ref{eq:stirringconf} (with a modulation period of $\tau = 30\ s$):

\begin{figure}[tbh]
 \centering
 \begin{center}
 \includegraphics[width=0.8\textwidth]{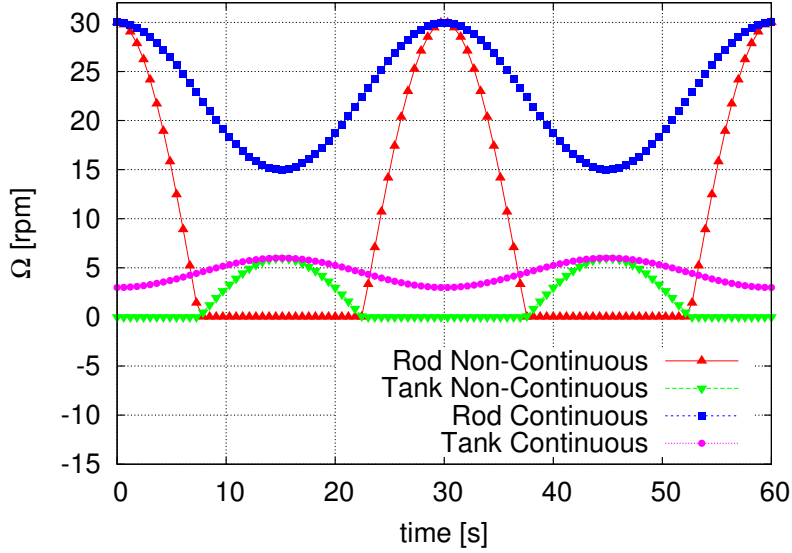}
\end{center}
 \caption{Temporal modulation of the angular velocity of both the rods and the tank for the continuous and non-continuous stirring cases. The modulation period, $\tau$, is $30 s$.}
 \label{fig:modulation}
\end{figure}

\begin{equation}
 \begin{array}{r c l l l}
  &&\text{Non-continuous case:}    & &\\
 \Omega_{1,2}&=&30 - 60 \sin^2(\frac{\pi t}{\tau})
	 &\text{if}\quad \Omega_{1,2}<0 & \Rightarrow \Omega_{1,2}=0\\
 \Omega_3 &=&6 - 12 \sin^2(\frac{\pi t}{\tau}+\frac{\pi}{2})
          &\text{if}\quad \Omega_{3}<0 & \Rightarrow \Omega_{3}=0\\
&&\text{Continuous case: \phantom{Non}}    & &\\
\Omega_{1,2} &=&30 - 15 \sin^2(\frac{\pi t}{\tau})& &\\
\Omega_3 &=&6 - 3 \sin^2(\frac{\pi t}{\tau}+\frac{\pi}{2})& &
 \end{array}
 \label{eq:stirringconf}
\end{equation}

The maximum angular wall velocities are fixed to $\Omega_1 = \Omega_2 = 30$ rpm for the rods and $\Omega_3 = 6$ rpm for the outer tank. Thus, the tangential wall velocity is the same and equal to $U = 31.41\ mm.s^{-1}$.
In our modeling, we take into account inertial effects because the flow that is considered for this tangential wall velocity does not satisfy the quasi-steady hypothesis.

\subsection{Rheological model}
\label{rheologicalmodel}

In order to model the flow of non-Newtonian fluids, we need to relate the shear stress tensor to the rate of deformation tensor. For generalized Newtonian fluids, the linear model, $\overset{=}{\tau}=2\:\eta\:\overset{=}D$, which has been established for purely-viscous fluids, is extended by replacing the constant viscosity by a function that depends on the shear rate. For Newtonian fluids, the relationship between the two tensors is simply linear, and the coefficient of proportionality is defined as the dynamic viscosity, $\eta$, of the fluid. The purely-viscous (i.e., inelastic) non-Newtonian character of the fluid that is studied here is represented by an Ostwald-de Waele power-law model for the case of both shear-thinning and shear-thickening fluids. The nonlinear relationship between the apparent viscosity, $\eta$, and the shear rate, $\dot{\gamma}$, is given by the constitutive equation:
\begin{equation}
    \eta\left(\dot{\gamma}\right) = k \left(\dot{\gamma}\right)^{n-1}
\end{equation}
with empirical constants $k$ and $n$, which are, respectively, the consistency index and the flow behavior index. The viscous stress tensor is then expressed as:
\begin{equation}
    \overset{=}{\tau}=2\eta\left(\dot{\gamma}\right)\overset{=}{D}
\end{equation}
where
\begin{equation}
\dot{\gamma} =  \sqrt{2\ tr\left(\overset{=}{D}^2\right)}
\end{equation}
and
\begin{equation}
\overset{=}{D} = \dfrac{1}{2}\left( \overset{=}{\nabla}V+(\overset{=}{\nabla}V)^T \right)
\end{equation}

 We will consider three different flow indexes that are respectively associated with shear-thinning, Newtonian and shear-thickening fluids: $n=0.5$, $n=1$ and $n=1.5$. The consistency index is adapted in each non-Newtonian case in order to the give the same generalized Reynolds number as was considered for the Newtonian flow (i.e., $Re = \dfrac{\rho\ U\cdot 2\cdot(R_3 - R_1)}{\mu} = 1.66$). This generalized Reynolds number, $Re_{PL}$, can be written for the power-law model as \cite{neofytou2005}:

\begin{equation}
    Re_{PL}= \frac {\rho \cdot (2(R_3 - R_1))^{n}} {k \cdot U^{n-2}}
\label{eq:REpl}
\end{equation}
thus, the consistency indexes are $k=0.939\ Pa.s^{0.5}$ for $n=0.5$ and $k=2.391\ Pa.s^{1.5}$ for $n=1.5$. The thermodependence of the viscosity and the viscous heating were not taken into account in this study. The Newtonian fluid that was considered in this study has the thermophysical properties that are listed in Tab.~\ref{tab:fluid}.

\begin{table}[tbh]
 \begin{center} 
\begin{tabular}{l l}\hline
Dynamic viscosity ($\mu$)& 1.5 $Pa.s$\\
Density ($\rho$) & 990 $kg.m^{-3}$\\
Thermal conductivity ($\lambda$) & 0.15 $W. m^{-1} K^{-1}$\\
Specific heat ($c_{p}$) & 1000 $W. kg^{-1} K^{-1}$\\
P\'eclet number ($Pe$) & $16,584$\\
\hline
 \end{tabular}  
\caption{Properties of the Newtonian fluid.\label{tab:fluid}}
\end{center}
\end{table}

\subsection{Thermal wall boundary condition}

Dirichlet boundary conditions are imposed on the rods and the tank walls. They are kept at a constant \emph{hot} temperature, $T_{hot}$. Before the start of the heating process, the initial uniform temperature of the fluid was set to a \emph{cold} temperature, $T_{cold}$. We define the dimensionless fluid temperature as:

\begin{equation}
 T^* = \dfrac{T-T_{cold}}{T_{hot} - T_{cold}}
 \label{eq:tempadim}
\end{equation}

Therefore, $T^*_{cold} = 0$ and $T^*_{hot}= 1$, and the maximum temperature difference between the walls and the fluid is always $1$. The heated walls play the role of a variable heat source that is continuously dissipated in the 2D modulated flow field. This is a specific situation that is not normally encountered when dealing with the scalar dissipation of a concentration field. The Dirichlet boundary condition, which is imposed on the walls, implies that during the mixing of the fluid in the tank the parietal fluxes will change along the walls and with time, depending on the local flow conditions.

The other important dimensionless parameter for this non-isothermal mixing problem is the Péclet number ($=Re.Pr$). The Péclet number can be seen as the ratio of the thermal diffusion time, $\tau_{td}$, to the advection time, $\tau_{ad}$, for scalar temperatures. The advantage of using the Péclet number, as opposed to the commonly used Prandtl number, is that the Péclet number is independent of the empirical power-law constants. Thus, its definition is valid for both Newtonian and non-Newtonian fluids. The limit Pe $= 0$ corresponds to the pure diffusion case. In our study, Pe is large (see Tab.~\ref{tab:fluid}) and hence, $\tau_{td}$ is $16,584$ times greater than $\tau_{ad}$. Thus, the need to speed-up the mixing of temperature scalars is clearly demonstrated.

\section{Mixing and energy indicators}
\label{mixingindicators}

In order to quantify the efficiency of the heating process for the Newtonian and non-Newtonian fluids, we have used two instantaneous measures: the mean value of the dimensionless temperature, $T^*_m$, and its standard deviation, $\sigma$. These quantities are defined as (where the summation is made over all of the mesh cells, $c$, with area $A_{c}$):
\begin{equation}
  T^*_m = \dfrac{1}{\sum_{c} A_{c}}\left(\sum_{c} A_{c} T^*_{c}\right) 
\label{eq:monitor1}
\end{equation} 
\begin{equation}
 \sigma = \left[\dfrac{1}{\sum_{c} A_{c}}\sum_{c}\left( A_{c}(T^*_{c} -T^*_m)^2\right) \right]^{\frac{1}{2}}
\label{eq:monitor2}
\end{equation}

The evolution of the mean temperature can be seen as an indicator of the total energy that is supplied to the fluid during the mixing process \cite{elomari2009a,elomari2008} while the standard deviation accounts for the homogenization level of the scalar temperature inside of the 2D tank.
The mean temperature is asymptotically bounded by the fixed temperature that is imposed on the walls (i.e., $T_{hot}$ or $1$ for non-dimensional temperatures). Efficient thermal mixing requires good values for both of the aforementioned indicators (i.e., $T^*$ near $1$ and $\sigma$ near $0$).

\section{Computational modeling}
\label{goveqnummeth}
\subsection{Governing equations}
The unsteady Navier-Stokes equations that govern the flow of incompressible fluids as well as the continuity equation are considered in their integral form:
\begin{multline}
 \dfrac{\partial}{\partial t} \int_{V}\rho\ \vec U\ \dd V
   + \int_{S} \rho\ \vec U \vec U \cdot \vec n \ \dd S =
   \int_{V} -\vec\nabla p \ \dd V  
+ \int_{S} \overset{=}{\tau}\cdot \vec n \ \dd S  \label{eq:mvt}
\end{multline}
\begin{equation}
 \int_{S} \vec U \cdot \vec n \ \dd S =0
\label{eq:4}
\end{equation} 
where $\overset{=}{\tau}$ is the viscous stress tensor, which is described in section \ref{rheologicalmodel} for both Newtonian and power-law fluids. The integration is over a volume, $V$, that is surrounded by a surface, $S$, whose orientation is described by its outward unit normal vector, $\vec n$. The energy conservation equation is considered in terms of the temperature:
\begin{eqnarray}
\dfrac{\partial}{\partial t} \int_{V}\rho c_p T\ \dd V
   + \int_{S} \rho c_p T \vec U \cdot \vec n \ \dd S &=&
   \int_{S} \lambda \vec\nabla T\cdot \vec n \ \dd S
\label{eq:3}
\end{eqnarray}
where $\lambda$ is the thermal conductivity.

\subsection{Numerical method}

The conservation equations (\ref{eq:mvt}, \ref{eq:4} and  \ref{eq:3}) are solved by means of an in-house code called Tamaris. This code has an unstructured, finite-volume framework that is applied to hybrid meshes. Variable values ($\vec U$, p and T) are stored at cell centers in a collocated arrangement. Cell shapes can be of different forms (tetrahedral, hexahedral, prismatic or pyramidal). 

Spatial schemes that approximate convective and diffusive fluxes are accurate to a second order. The convective fluxes are approximated by the non-linear high-resolution bounded scheme, CUBISTA, of Alves \textit{et al.} \cite{alves2003} and by using the deferred-correction practice of Ng \textit{et al.} \cite{ng2007}. Diffusive terms are treated with a centered-differencing scheme in conjunction with a treatment of possible non-orthogonality of the mesh \cite{jasak1996}. 

Pressure-velocity coupling is ensured by the SIMPLE algorithm \cite{patankar1980} while the mass fluxes at the cell faces are evaluated with the Rhie-Chow interpolation \cite{rhie1983} in order to avoid pressure checkerboarding. The implicit three-time-step Gear's scheme with a second-order accuracy is used to discretize the unsteady terms at each iteration; the discretization technique that was presented above leads to a linear system with a non-symmetric sparse matrix for each variable. These linear systems are solved by means of an ILU preconditioned GMRES solver using the implementation of the IML++ library \cite{dongarra1994}.

 This 3D code can deal with 2D computations (e.g., in ($\vec x,\vec y$) plane), without any change, by considering a single layer of computational cells (in $\vec z$ direction) and by neutralizing the top and bottom faces (with respect to $\vec z$). In the scope of this work, all of the computational meshes were generated with the open-source software Gmsh \cite{geuzaine2009}.

\subsection{Validation test cases}

\begin{figure}[htbp]
\centering
\begin{center}
\includegraphics[width=0.5\textwidth]{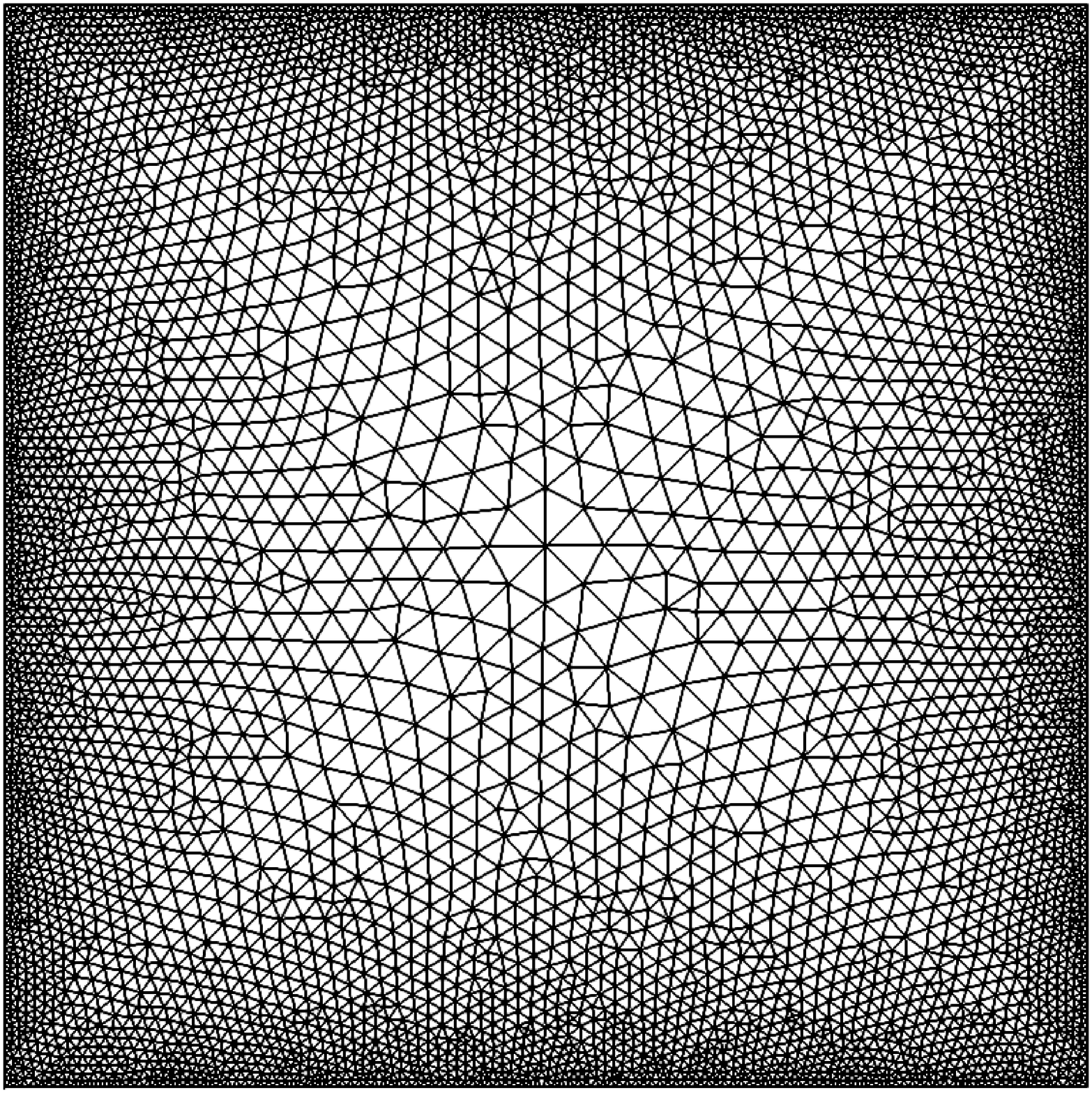}
\end{center}
\caption{Computational mesh for the lid-driven square cavity mesh ($10468$ cells). }
\label{fig:lidcav}
\end{figure}

\begin{figure}[htbp]
\centering
\begin{center}
\includegraphics[width=0.5\textwidth]{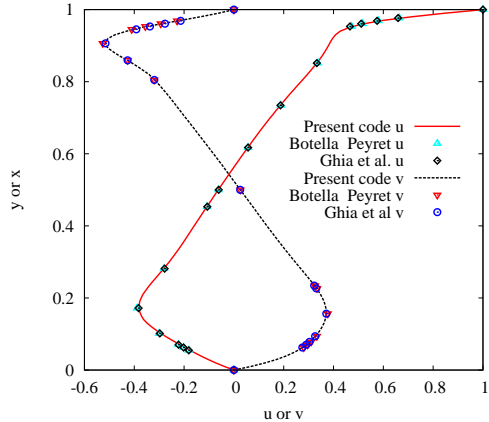}
\end{center}
\caption{For the Newtonian fluid, the $u$ and $v$ velocity components are presented for the lid-driven cavity test case ($Re=1000$).}
\label{fig:lidcavNt}
\end{figure}

In this section, we perform calculations for some well-documented test cases in order to assess the accuracy of the code. We consider the flow in the lid-driven, square cavity for three different fluids: Newtonian, shear-thinning (with $n=0.5$) and shear-thickening (with $n=1.5$). The mesh that is used in each of the three cases (Fig. \ref{fig:lidcav}) is composed of $10,468$ triangular cells. For the Newtonian fluid, we present the results for the case with a Reynolds number of $Re=1,000$. The results are then compared to the results of Botella and Peyret \cite{botella1998}, which were obtained by using a spectral Chebyshev method with a polynomial of degree $N=96$, and to those of Ghia \textit{et al.} \cite{ghia82}, which were obtained by using a vorticity-stream function formulation and a $128\times 128$ grid. In Fig. \ref{fig:lidcavNt}, the velocity components $u$ and $v$ are both presented along the vertical ($y$) and horizontal ($x$) cavity axes, respectively. By comparing the extrema values of  velocity components $u$ and $v$ and their locations to those obtained by Botella and Peyret \cite{botella1998} we found relative differences lower than $2\%$.

\begin{figure}[htbp]
\centering
\begin{center}
\includegraphics[width=0.5\textwidth]{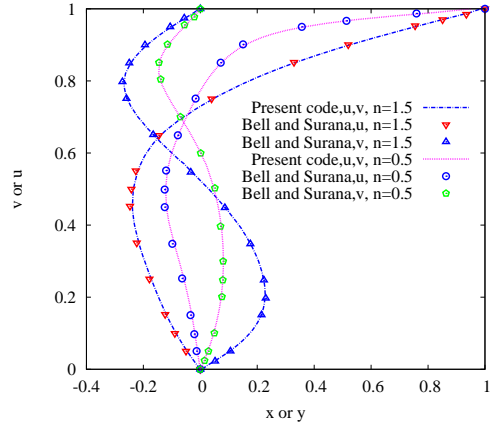}
\end{center}
\caption{For the shear-thinning and shear-thickening fluids, the $u$ and $v$ velocity components are presented for the lid-driven cavity test case ($Re_{PL}=100$).}
\label{fig:lidcavNN}
\end{figure}

The code is also validated by comparing the results of the non-Newtonian power-law fluids to the results of Bell and Surana \cite{bell1994}, which were obtained by using a p-version least-squares finite-element method. The flow that is considered has a power-law Reynolds number of $Re_{PL}=100$ (Eq. \ref{eq:REpl}) for both values of $n$. The results are shown in Fig. \ref{fig:lidcavNN}. For the two fluids that were studied, the results that are given by our code compare quite well with those of the other authors. The relative difference between the results of the extrema of velocity and of their locations are less than $2.5\%$.

\section{Results and discussion}
\label{results}
\begin{figure}[htbp]
 \centering
 \begin{center}
 \includegraphics[width=0.5\textwidth]{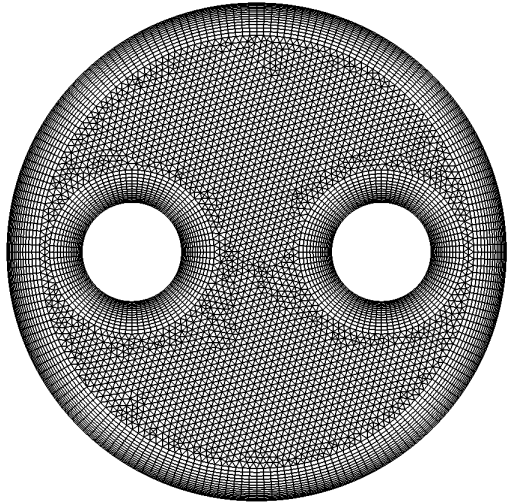}
\end{center}
 \caption{Computational mesh of the two-rod mixer.}
 \label{fig:mixer_mesh}
\end{figure}

After a grid size-dependence study (Appendix A), a mesh of $10,680$ computational cells is adopted. This mesh is shown in Fig. \ref{fig:mixer_mesh}, where regular quadrilateral cells are used near the walls to enhance the resolution of the boundary layers. 

\subsection{Temporal evolutions of the temperature statistics indicators}\label{sec:statistics}

\begin{figure}[htbp]
 \centering
 \begin{center}
 (\textbf{a}) \includegraphics[width=0.6\textwidth]{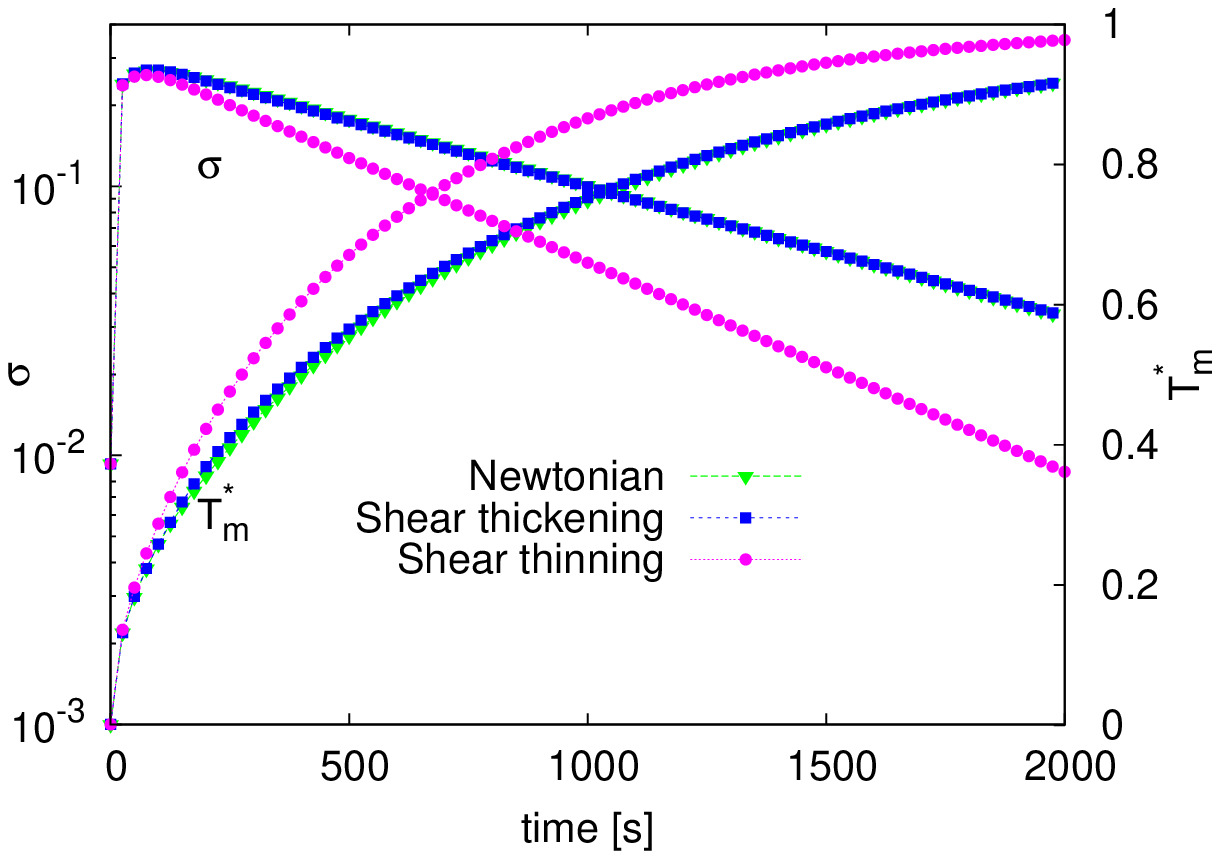}\\
 (\textbf{b}) \includegraphics[width=0.6\textwidth]{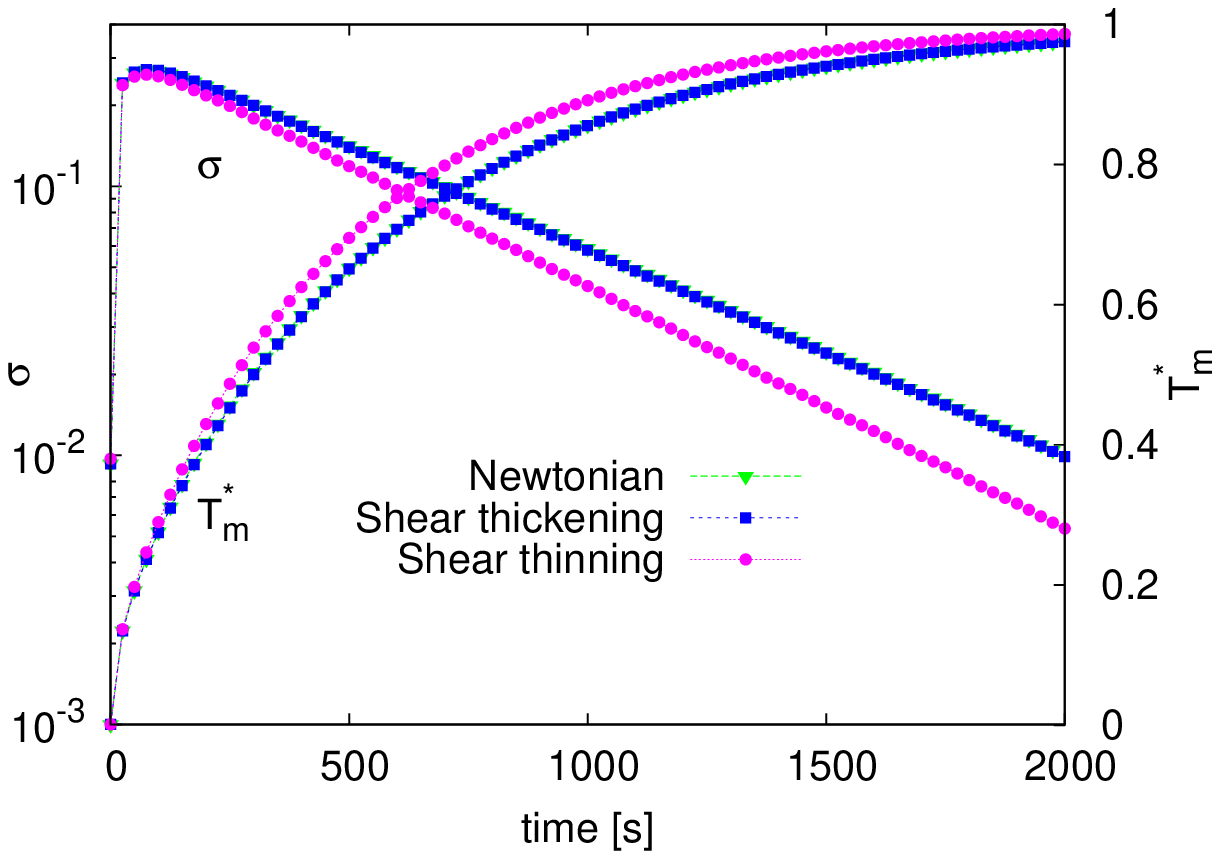}\\
 (\textbf{c}) \includegraphics[width=0.6\textwidth]{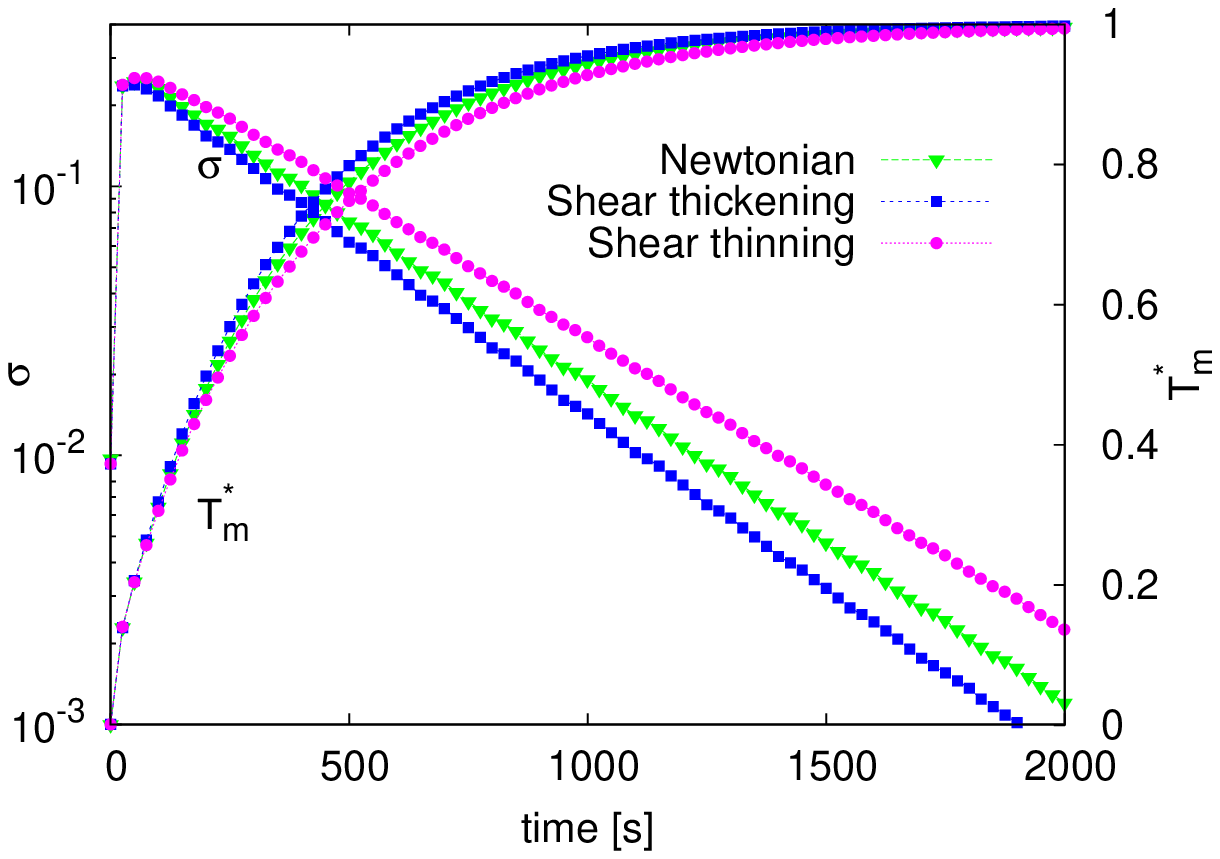}
\end{center}
 \caption{For the three rheological fluid behaviors, the temporal evolution of the mean value of the dimensionless temperature and its standard deviation is presented for (a) non-modulated, (b) continuously-modulated and (c) non-continuously-modulated stirring protocols.}
 \label{fig:SD_Tm}
\end{figure}

For the three rheological fluid behaviors, the temporal evolutions of the mean value of the dimensionless temperature, $T^*_m$, and its standard deviation, $\sigma$, are shown in Fig. \ref{fig:SD_Tm} for the following stirring protocols: (a) non-modulated, (b) continuously modulated and (c) alternately modulated. The fixed-wall temperature that is imposed on all of the  boundaries acts as a source for the evolution of the scalar; thus, the mean temperature will not be constant with time but will evolve asymptotically from $T^*_m = 0$ to $T^*_m = 1$. 
As can be noticed from the curves of $\sigma$, its decay is exponential. Nevertheless the decay rate is low for the non modulated stirring (Fig. \ref{fig:SD_Tm}(a)) and is the strongest for the alternately modulated stirring (Fig. \ref{fig:SD_Tm}(c)), and this whatever the rheology of the fluid. In the case of stirring without modulation, large areas of cold fluid are isolated between the rods and the tank because the velocity field is constant over time and the flow patterns are stationary which do not promote mixing. It is a non-chaotic (regular)  flow. For the continuously modulated protocol (Fig. \ref{fig:SD_Tm}(b)), these areas still exist but are smaller since some time variation has been introduced into the flow, which becomes partially chaotic (i.e., chaotic and regular regions coexists). This results in an increase of the decay rate. When a non-continuous modulation of the wall velocity is considered (Fig. \ref{fig:SD_Tm}(c)), the exponential decay rates of the standard deviation of the temperature fields are always higher than for the two other types of stirring protocols that have been studied (Fig. \ref{fig:SD_Tm} (a)  and (b)).  With alternated modulation,  the Newtonian and shear-thickening fluids do not present unmixed zones: it is a globally chaotic flow. The shear-thinning fluid presents small unmixed zones at the beginning of the mixing process, which explains its decay rate weaker than those of the other two fluids.

The behavior of both the Newtonian and shear-thickening fluids is exactly the same for the non-modulated and continuously-modulated stirring protocols. They both give a poorer mixing efficiency than the shear-thinning fluid (Fig. \ref{fig:SD_Tm}(a) and (b)). However, this situation is reversed for the non-continuously-modulated stirring protocol for which the shear-thickening fluid gives a more efficient thermal mixing.  These observations are corroborated by the study of the probability distribution functions of the dimensionless temperature, $T^*_m$ in section \ref{sec:pdf} and would be best explained by the examination of the velocity fields presented at the next section. In Tab. \ref{tab:sig2000} we summarize the values of the standard deviation reached after $2000 s$ of mixing for the three fluids and the three stirring protocols. 

 \begin{table}[th]
 \begin{center} 
\begin{tabular}{l c c c}\hline
         &  NM   & CM  &  ALT \\
  \hline
Shear-thinning   &  $8.63\times 10^{-3}$  & $5.34\times 10^{-3}$  & $2.24\times 10^{-3}$\\
Newtonian         &   $3.26\times 10^{-2}$ &  $1.00\times 10^{-2}$ & $1.20\times 10^{-3}$\\
Shear-thickening &   $3.29\times 10^{-2}$  & $0.98\times 10^{-2}$  & $7.64\times 10^{-4}$ \\
\hline
 \end{tabular}  
\caption{The values of $\sigma$ reached by the three fluids at $t=2000 s$ for the non-modulated (NM), continuously-modulated (CM) and alternated (ALT) stirring protocols. \label{tab:sig2000}}
\end{center}
\end{table}

\subsection{Non-Newtonian flow patterns}

\begin{figure}[ptbh]
 \centering
 \begin{center}
\includegraphics[width=0.9\textwidth]{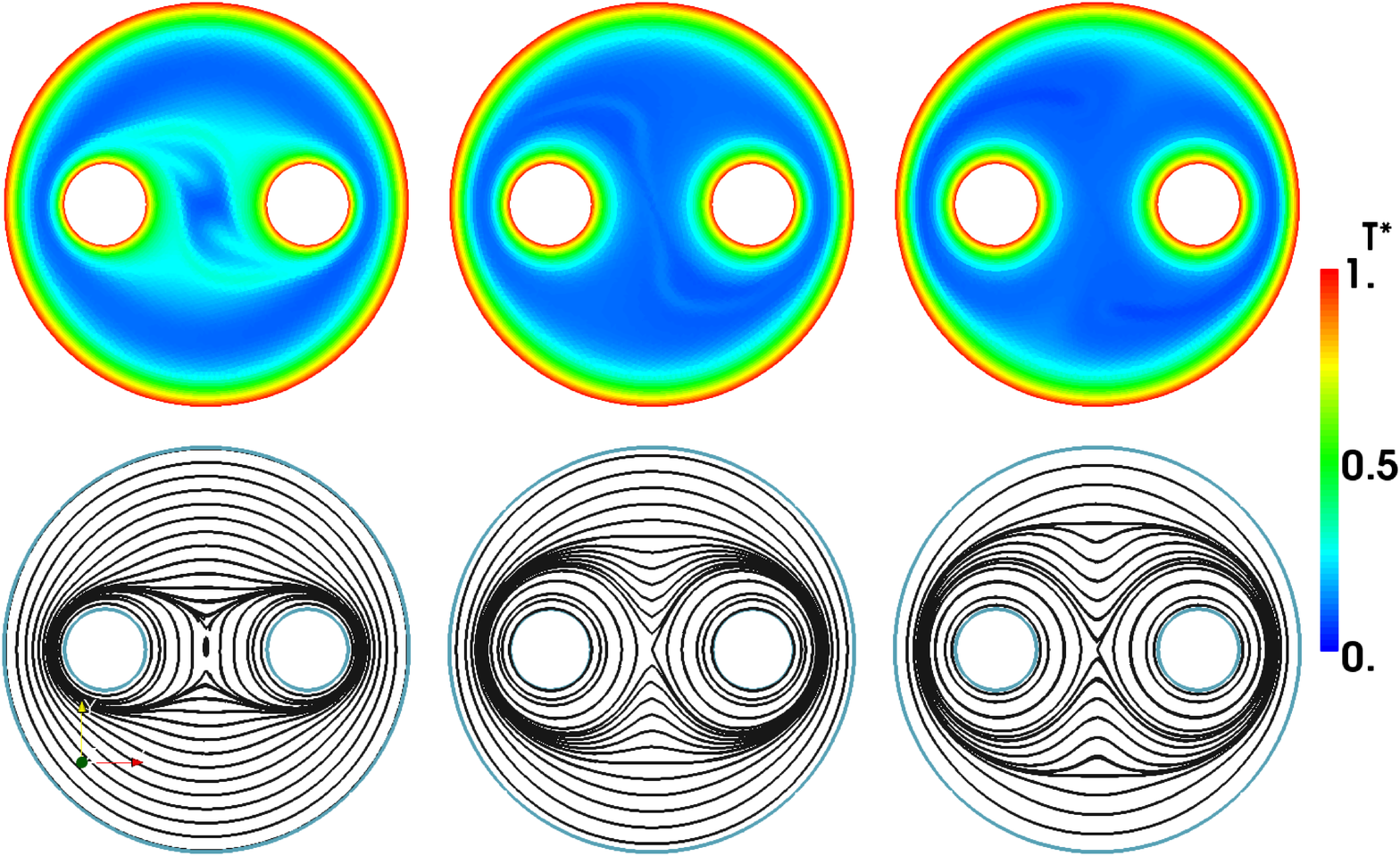}\\$t = 120\, s = 4\ \tau$ \rule{\textwidth}{1pt}
\includegraphics[width=0.9\textwidth]{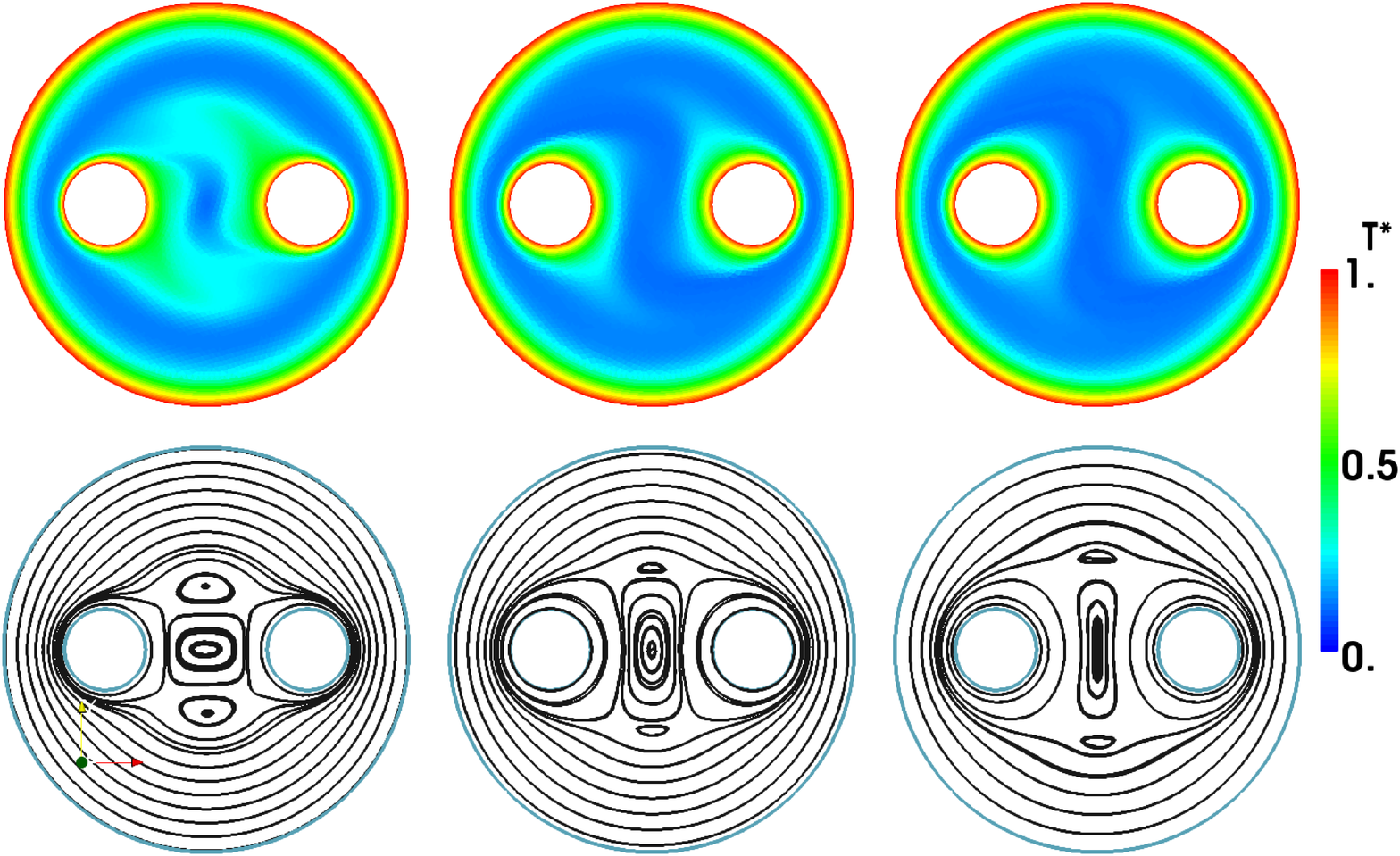}\\$t = 135\, s = 4\ \tau + 1/2\ \tau$
\end{center}
 \caption{For the case of continuously-modulated stirring, the flow patterns (temperature fields and streamlines) at two different instants ($t = 120\ s$, top half, and $t = 135\ s$, bottom half) are presented for the three rheological fluid behaviors: shear-thinning fluid (left), Newtonian fluid (middle) and shear-thickening fluid (right). The period of modulation is $\tau = 30 s$.}
 \label{fig:Tiso_strm_MC}
\end{figure}

\begin{figure}[ptbh]
 \centering
 \begin{center}
\includegraphics[width=0.9\textwidth]{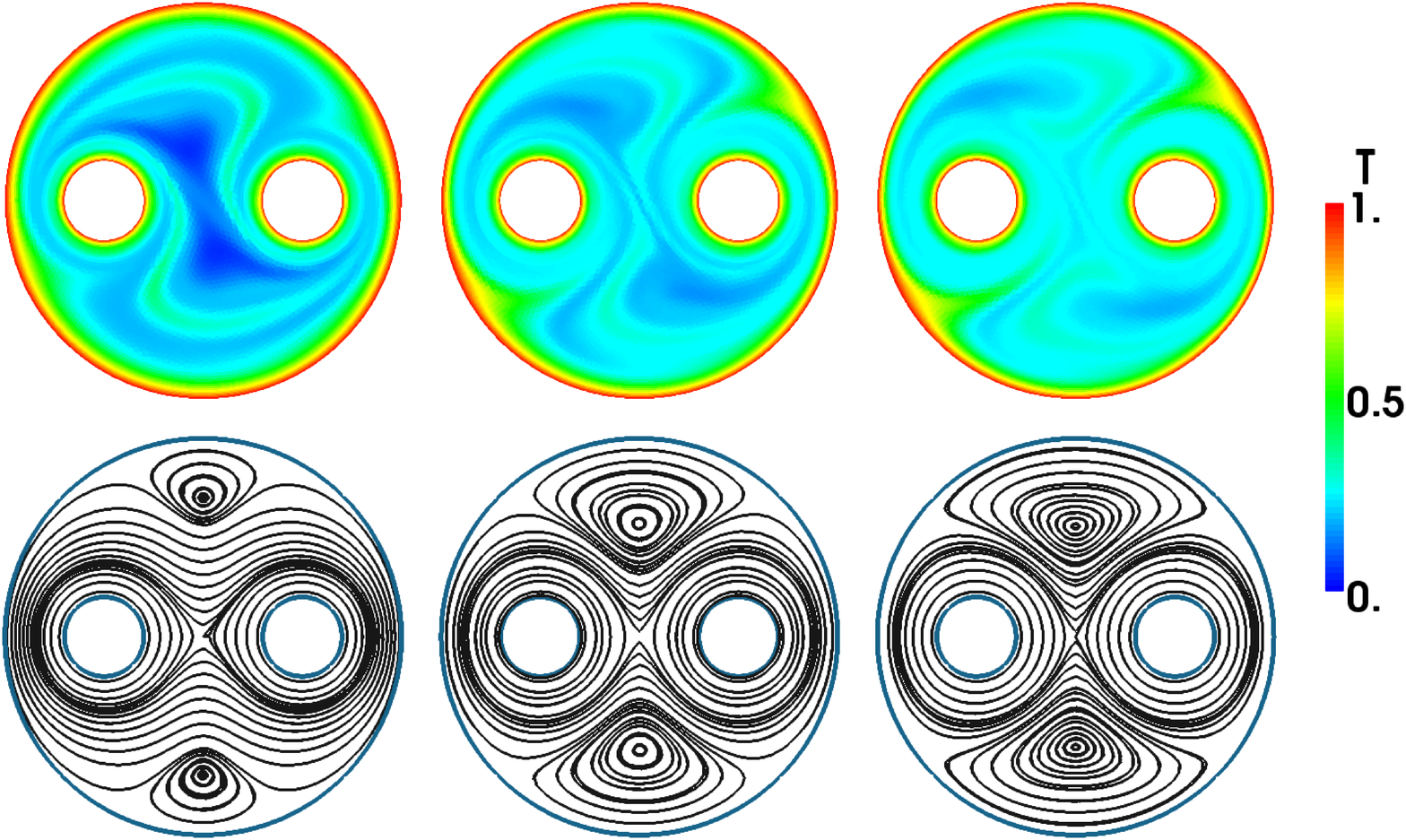}\\$t = 120\, s = 4\ \tau$ \rule{\textwidth}{1pt}
\includegraphics[width=0.9\textwidth]{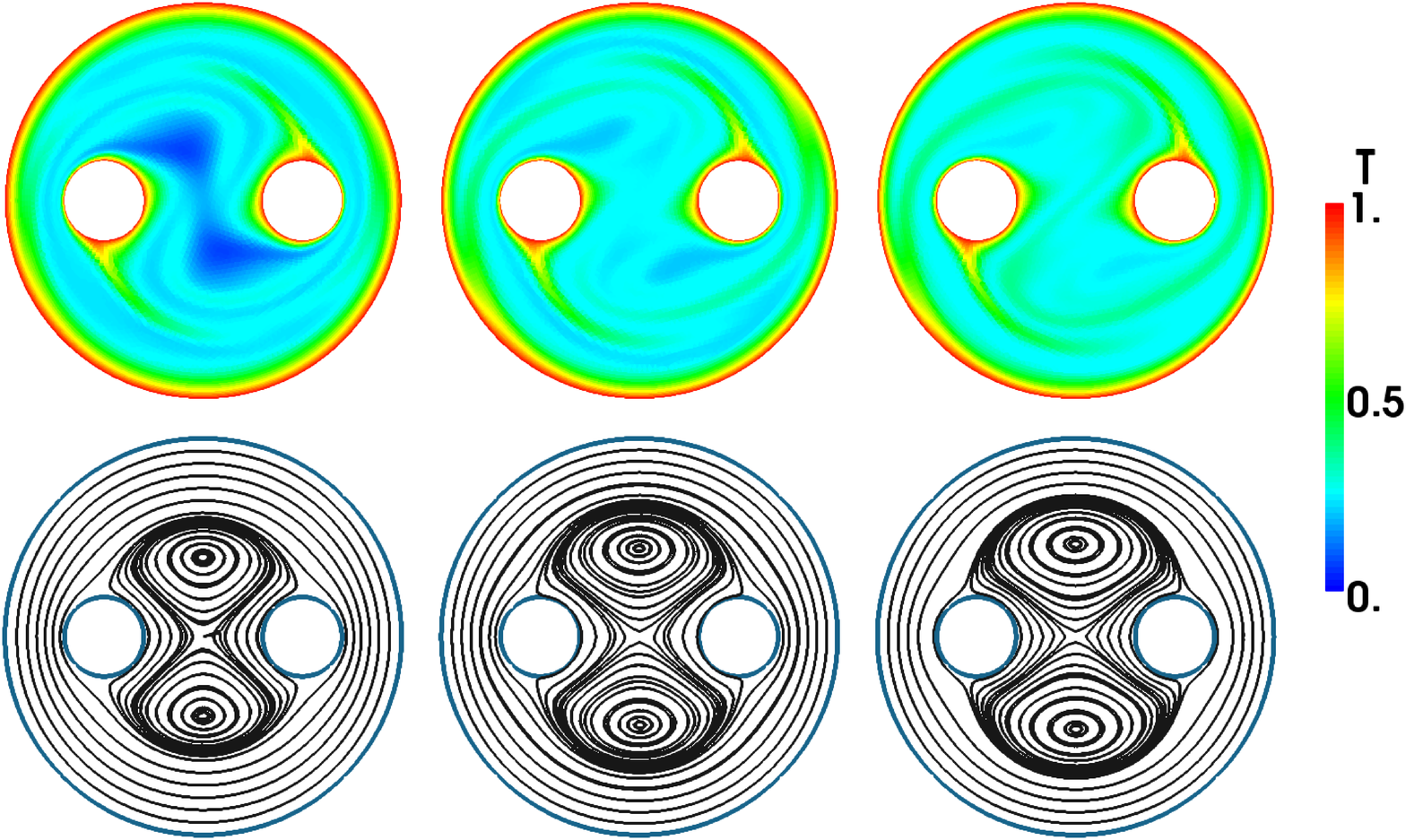}\\$t = 135\, s = 4\ \tau + 1/2\ \tau$
\end{center}
 \caption{For the case of alternately-modulated stirring, the flow patterns (temperature fields and streamlines) at two different instants ($t = 120\ s$, top half, and $t = 135\ s$, bottom half) are presented for the three rheological fluid behaviors: shear-thinning fluid (left), Newtonian fluid (middle) and shear-thickening fluid (right). The period of modulation is $\tau = 30 s$.}
 \label{fig:Tiso_strm_ALT}
\end{figure}

In order to compare the flow topologies that have been obtained for the three different rheological behaviors, we present in Figs.  \ref{fig:Tiso_strm_MC} and \ref{fig:Tiso_strm_ALT} the patterns of the scalar temperature fields that are associated with the corresponding streamlines for both continuously- and non-continuously-modulated stirring protocols.

In the first case (continuous modulation), the upper part of Fig. \ref{fig:Tiso_strm_MC} represents the flow situation at time $t=120\ s$, which is equivalent to $t=4 \tau$, while the lower part corresponds to the instant $t=135\ s = 4.5 \tau$. At $t=120\ s$, the two rods are rotating at their maximum angular velocity while the velocity of the tank is at its minimum value (see Fig. \ref{fig:modulation}). By examining the streamlines, we notice that the size of the zone of influence of the rods, \textit{i.e.} the zone of the driven fluid, depends on the fluid's nature: it is confined for the case of the shear-thinning fluid and is larger for both the Newtonian and the shear-thickening fluids. For the latter, this zone is the largest. The effect on the temperature field is the opposite: for Newtonian and shear-thickening fluids, the streamlines that are parallel to the surface of the rods form large zones and act as insulation, which restrains the convective heat transfer from the rods. On the other hand, in the case of the shear-thinning fluid, the narrowness of this zone induces fluid streams between the surfaces of the two rods and promotes heat transfer from them to the center of the mixer.

Later at $t=135\ s$, the tank is rotating at its maximum velocity while the rods are at their minimum. A fluid recirculation zone appears between the rods. It is larger when the fluid is less viscous because the rods have narrower zones of driven fluid. Thus, for the shear-thinning fluid, this recirculation blends the heated fluid, which was extracted from the rods earlier. As it was observed by statistical analysis of Fig. \ref{fig:SD_Tm}(b), the best mixing results are achieved for this fluid when the stirring is continuously-modulated ($\sigma = 5.34\times 10^{-3}$ at $t=2000 s$, while $\sigma = 1\times 10^{-2}$ for Newtonian and shear-thickening fluids, see Tab. \ref{tab:sig2000}).
 
 Contrary to Newtonian and shear-thickening fluids, the shear-thinning fluid mixing benefits from the inertial effects due to the continuous rotation in the cases of non modulated (not showed here) and continuously modulated protocols. These effects result from the decrease of the viscosity and the creation of a significant fluid recirculation between the two rods. Because of higher viscosities, the two other fluids do not present the same mechanism and the streamlines around the rods are closed.

Newtonian and shear-thickening fluids exhibit different velocity fields nevertheless they are sufficiently comparable to result in quite similar temperature fields, that give the same global values $T^*_m$ and $\sigma$ as it was observed in Figs. \ref{fig:SD_Tm}(a) and (b).  This is most probably due to the relatively high thermal diffusivity of the fluids (compared to molecular diffusivity) that cancels the small differences between the temperature patterns of the two fluids. Moreover, this high similarity between the behaviors of the two fluids in the case of continuous modulation can be related to the fact that the efficiency of the thermal mixing in this mixer depends on the heat transfer at the heated boundaries. Even if the velocity fields of the two fluids present some discrepancies in the center of the mixer, in the vicinity of the walls they both present almost parallel streamlines. Shear-thinning fluid presents different flow topology near the surfaces of the rods which results in lower values of standard deviation.

During all this mixing process the streamlines are parallel to the tank surface and as a consequence, hot fluid advection towards the center of the mixer is very weak. This feature is more pronounced for the non-modulated stirring case.

In the case of alternating rotations, Figure \ref{fig:Tiso_strm_ALT} is the equivalent of Fig. \ref{fig:Tiso_strm_MC}. Hence, the upper part of the figure ($t=4 \tau$) corresponds to the tank at rest for $\tau/2$ while the rods are rotating. By examining the streamlines, we can observe that the more the fluid is viscous the more the effect of the rods extends towards the tank to form four parabolic points on its boundary. Also, two symmetrical vortices are created near this boundary, their size is very small for the shear-thinning fluid. As a consequence, these vortices cause two hot fluid streams to be extracted from the tank wall and have their origin at two of the four parabolic points. In the case of the shear-thickening fluid, these streams almost reach the center of the tank. During this period we can note that there is no parallel streamlines to the tank.

At $t=4.5 \tau$ (the lower part of the Fig. \ref{fig:Tiso_strm_ALT}), the tank rotation is at its maximum velocity while the rods are at rest. We observe two recirculation zones in the center of the tank with a size that increases with increasing fluid viscosity. Moreover, two parabolic points form on each rod and hot fluid streams escape from these points. We can clearly see that the temperature field contains unmixed cold zones for the case of the shear-thinning fluid and that the best homogenization is obtained for the shear-thickening fluid. In comparison with Fig. \ref{fig:Tiso_strm_MC}, we can observe that the alternated modulation produces a hotter fluid and gives a better homogenization for the three fluids. These visual observations confirm the statistical results of Fig. \ref{fig:SD_Tm}.

It is during the period of rotation of the rods that the three fluid behaviors exhibit essential differences.  The more the fluid is viscous, the more the four parabolic points at the tank surface are distant from each other and the more the separatrices that emanate from these points are extended.  The fluid driven along these separatrices causes more heat extraction from the tank wall.  The mixing mechanism of the alternated protocol is a succession of stretching and folding of hot fluid striations that is made possible by the crossing of the streamlines between first half of the modulation period to the second half. The fluid stretching is more efficient for viscous fluids, for which the mixing exhibits a chaotic character.  The shear-thinning fluid develops thin boundary layers around the rods when they rotate, as a consequence, little fluid stretching is achieved and unmixed cold zones appear in the mixer. Contrarily to the two other fluids, the mixing of the shear-thinning fluid is not globally chaotic.  In the alternated modulation case, the differences in the flow topologies near the tank boundary are more pronounced between shear-thickening and Newtonian fluids than those observed in the continuously-modulated protocol (Fig.  \ref{fig:Tiso_strm_MC}), leading to contrasting temperature results (Fig. \ref{fig:SD_Tm}).

From all the results, we can deduce that full chaotic mixing is achieved for the alternated protocol and when highly viscous fluid is concerned.

\subsection{The effect of the modulation's period size}

\begin{figure}[htbp]
 \centering
 \begin{center}
 \includegraphics[width=0.48\textwidth]{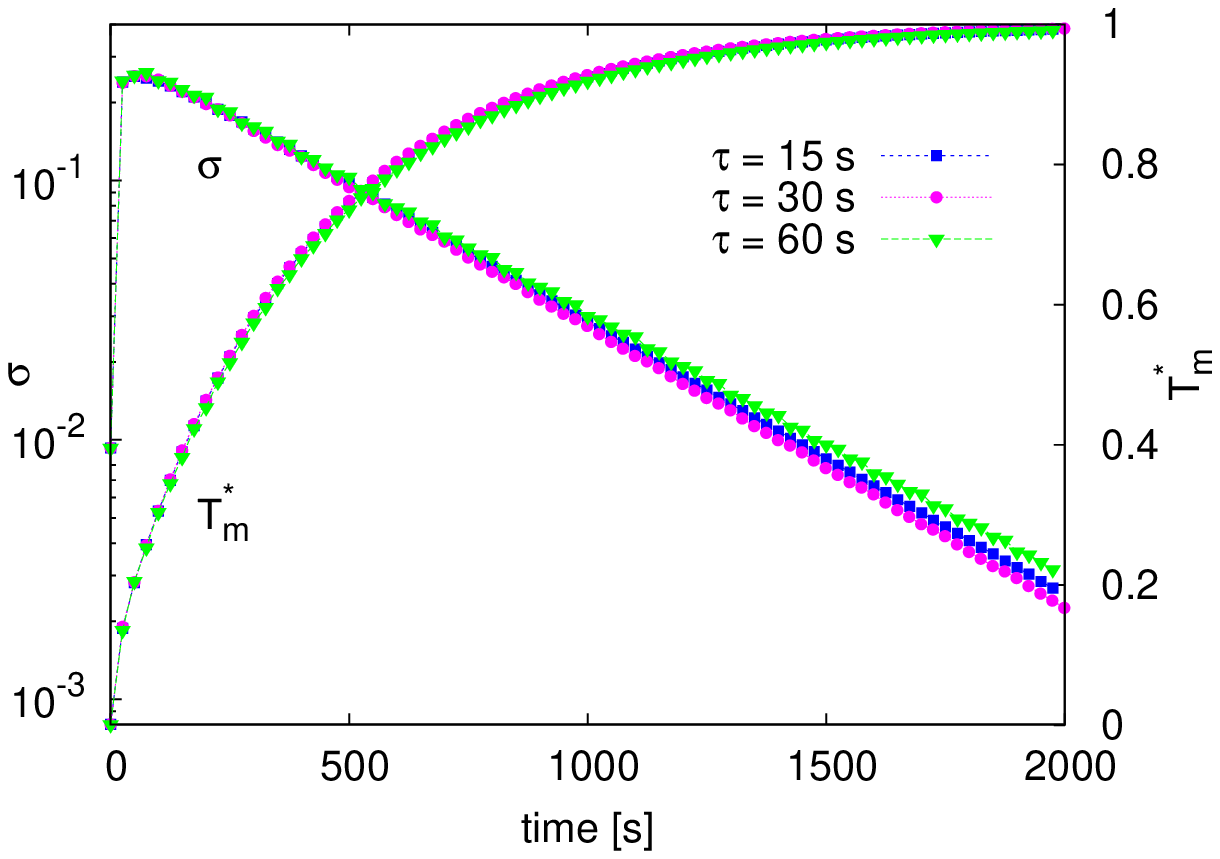} 
 \includegraphics[width=0.48\textwidth]{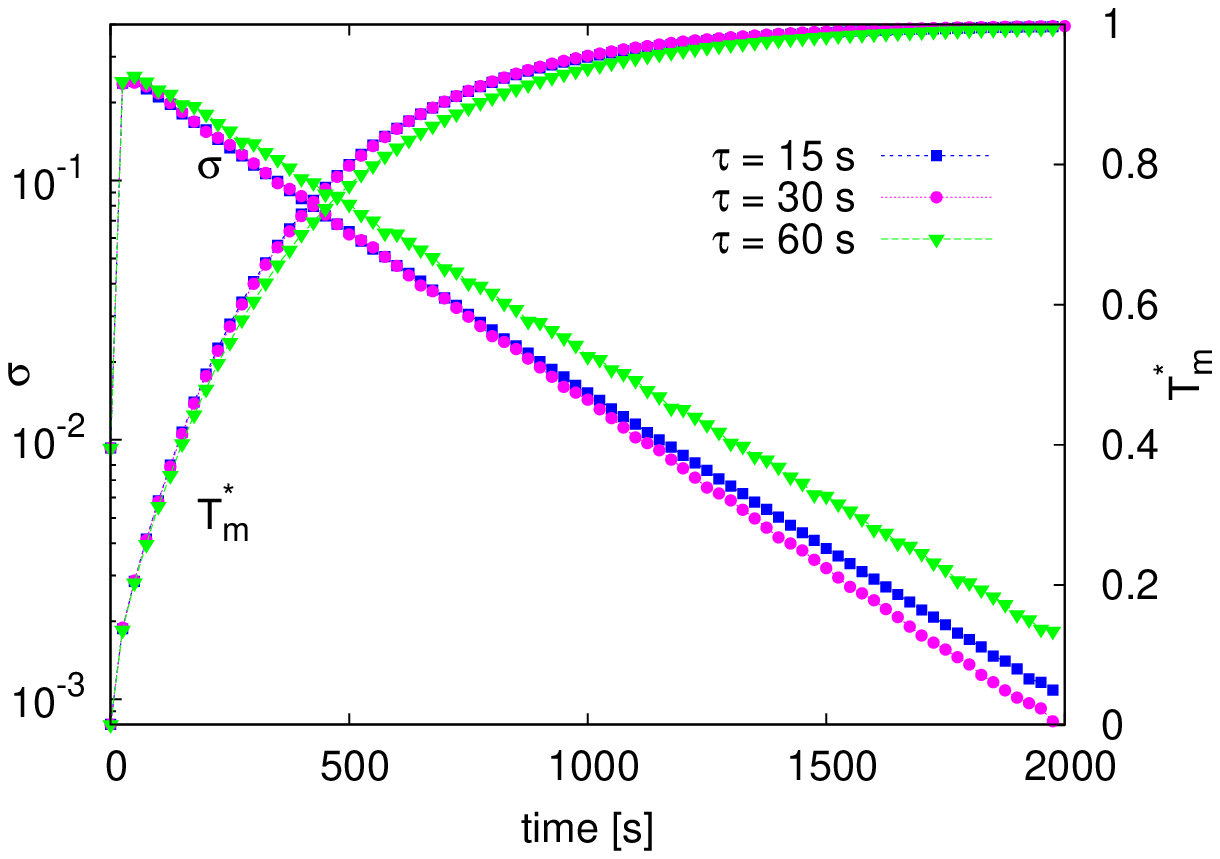} 
\end{center}
 \caption{Temporal evolution of the mean value of the dimensionless temperature and its standard deviation for three different period sizes. Left: shear-thinning fluid, right: shear-thickening fluid.}
 \label{fig:SD_Tm_Per}
\end{figure}

Until now the period size was fixed at the value $\tau = 30\ s$. In this section, we investigate the effect of increasing or decreasing this period size in the case of alternating modulation. For both shear-thinning and shear-thickening fluids, Fig. \ref{fig:SD_Tm_Per} presents the results that have been obtained for mixing processes with periods of $15\ s, 30\ s$ and $60\ s$. 

For the two fluids, the size of the period has a slight effect on the evolution of the mean temperature. For the standard deviation, the mixing of the shear-thickening fluid is impacted by the change of this parameter while the mixing of the shear-thinning fluid presents little difference. For both fluids, the period size of $\tau=30\ s$ seems to give the best results.  
In our previous studies \cite{elomari2009a,elomari2009b}, we have found similar conclusions for Newtonian fluids. An explanation can be advanced by observing the evolution of the temperature field in the mixer (as the one in Fig. \ref{fig:Tiso_strm_ALT}) during several periods of the mixing process. It is found that when the period is shorter ($\tau=15\ s$) the hot fluid streams that are extracted from the walls are smaller; thus, they are returned back to the boundary during the next movement of the wall. On the other hand, when the period is longer ($\tau=60\ s$), the mixing process loses some of its temporal diversification or its unsteady character because a pseudo-steady state is reached during each of the wall's movements. The value $\tau=30\ s$ is not the optimal value but we think that it is around it.

\subsection{Probability distribution functions of the temperature scalars}\label{sec:pdf}

\begin{figure*}[htbp]
 \centering
 \begin{center}
 \includegraphics[width=0.45\textwidth]{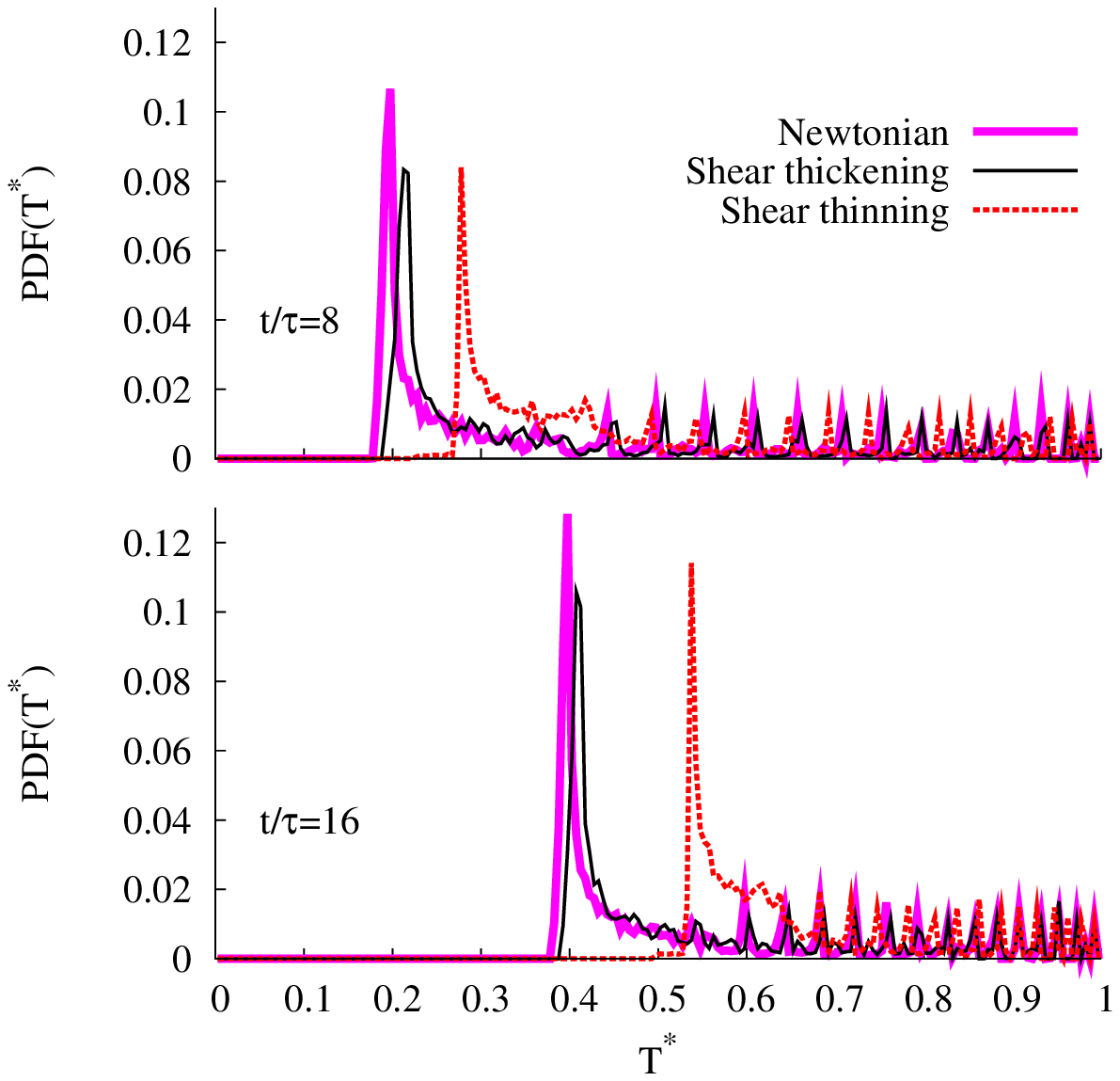} 
  \includegraphics[width=0.45\textwidth]{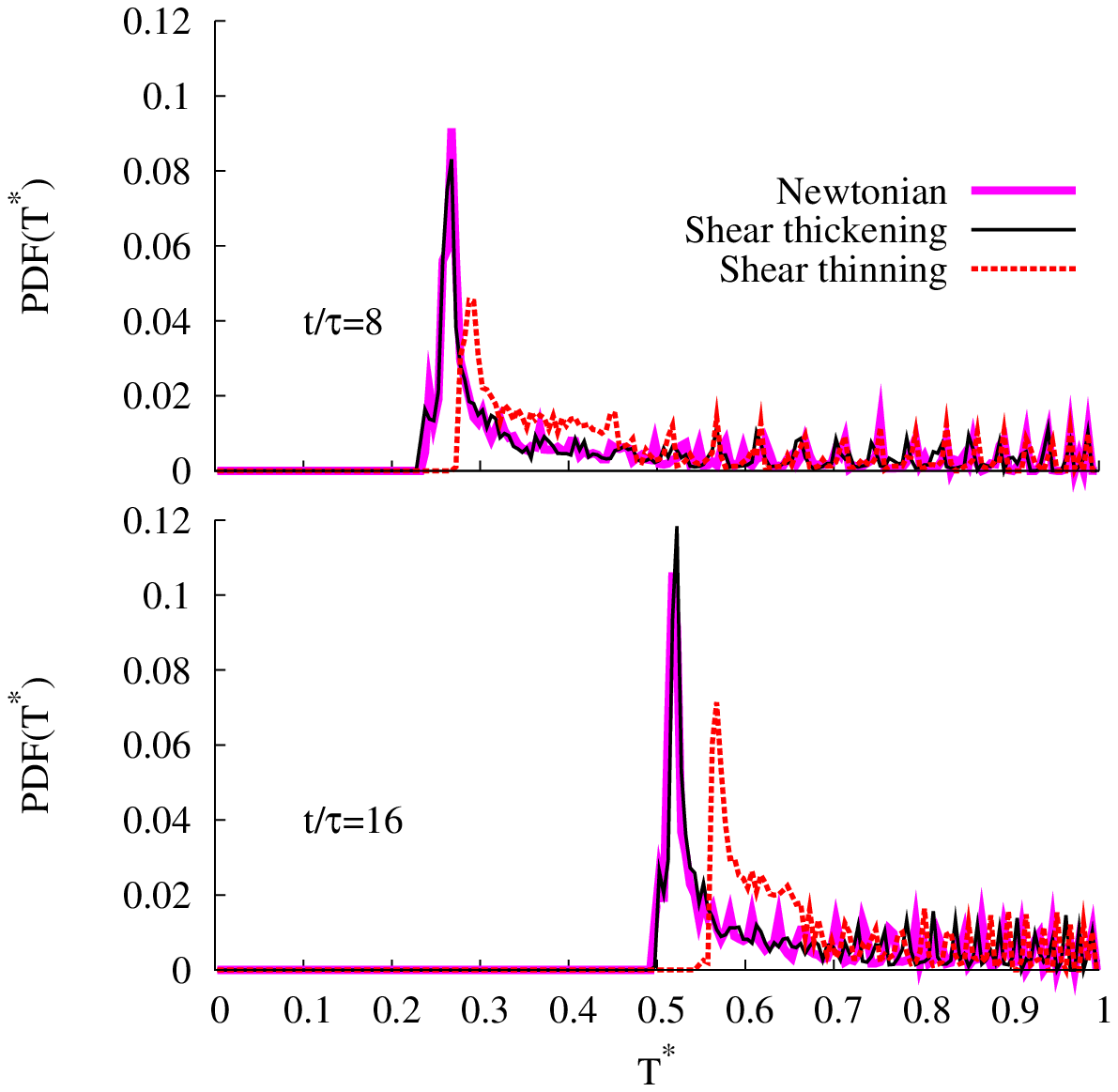}\\
(\textbf{a}) \hspace*{0.33\textwidth}(\textbf{b})\\
   \includegraphics[width=0.45\textwidth]{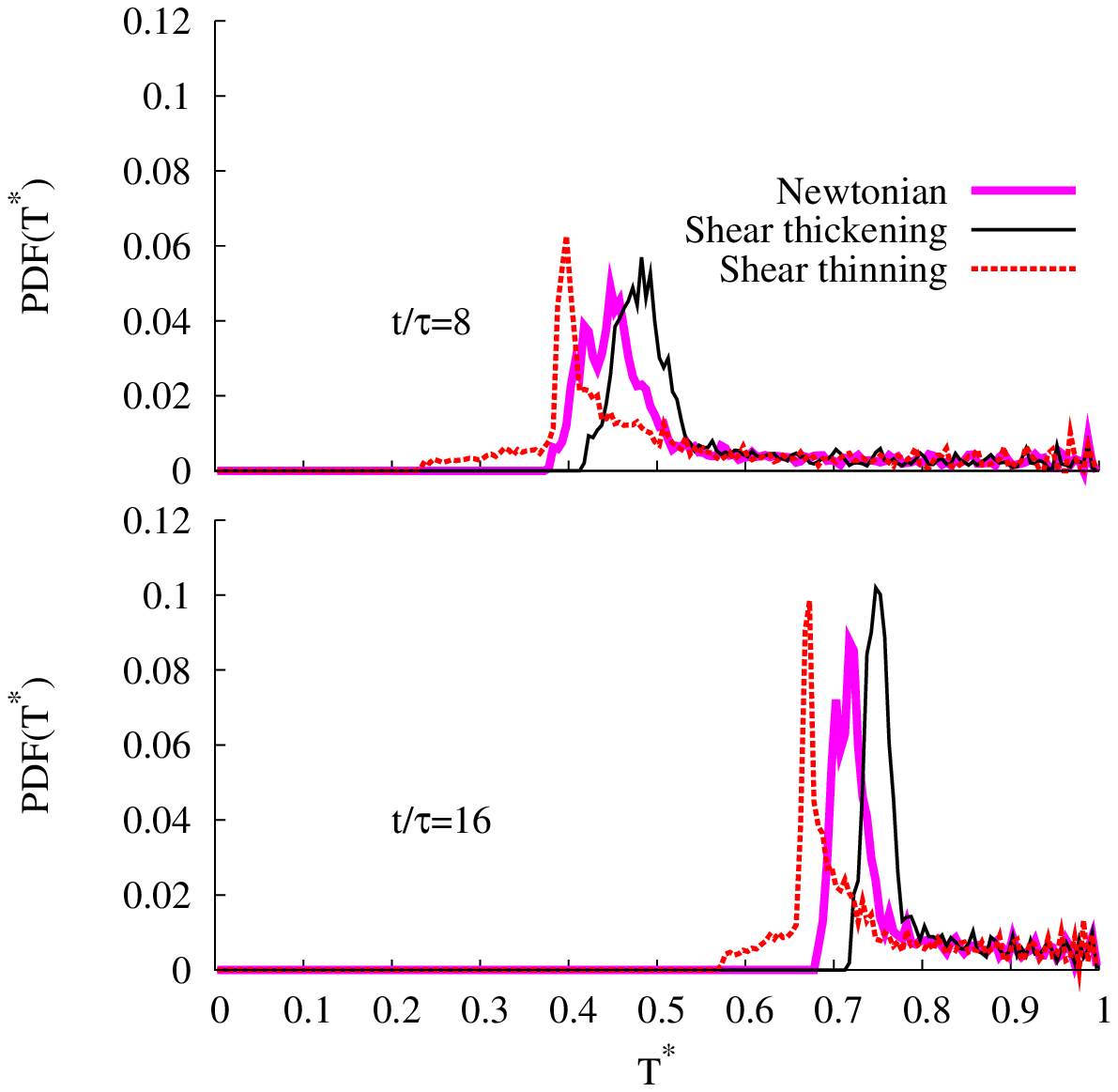}\\(\textbf{c}) 
\end{center}
 \caption{Probability distribution functions (PDFs) of $T^*$ for the three rheological fluid behaviors at two different times and for the (a) non-modulated, (b) continuously-modulated and (c) non-continuously-modulated stirring protocols.}
 \label{fig:PDF_T}
\end{figure*}

We focus here on the probability distribution functions (PDFs) of the dimensionless temperature, $T^*$, for the whole mixer section that is filled by the fluid. The PDFs of $T^*$ fields are shown in Fig. \ref{fig:PDF_T} for the three stirring protocols at two different times. In each plot, the three fluid behaviors are compared. 

In general and in all of the figures, we notice the presence of a significant peak (which corresponds to the most probable temperature in the fluid) that is located on the left. To the right of this peak, for the non-modulated stirring protocol (Fig. \ref{fig:PDF_T} (a)), we observe some small modulations that indicate the persistence of the scalar temperature gradients within the fluid. Then, with time, as the peak moves towards the high temperatures, these modulations compress towards $T^* = 1$. In this case, the shear-thinning fluid gives the best thermal mixing efficiency; thus, the results that have been found in section \ref{sec:statistics} are then confirmed. For the continuously-modulated stirring protocol (Fig. \ref{fig:PDF_T}(b)), we observe the same general features as in Fig. \ref{fig:PDF_T} (a); the PDFs are just translated a little towards $T^* = 1$. The difference is now considerably reduced between the $T^*$ location of the PDF peak that is observed in shear-thinning fluids and the two other fluids. The PDFs that correspond to the non-continuously-modulated stirring protocol (Fig. \ref{fig:PDF_T} (c)) are completely different; the peak of the most probable temperature is wider and its tail is reduced towards $T^* = 1$ and has smaller modulations. Contrary to the results that have been obtained for the two other stirring protocols, the shear-thinning fluid gives the poorest thermal mixing efficiency. In this case, the $T^*$ peak conserves a left tail towards the cold temperatures, which indicates the persistence of cold, poorly-mixed zones within the fluid; these zones can be seen in Fig. \ref{fig:Tiso_strm_ALT} (left). With Fig. \ref{fig:PDF_T} (c), we can conclude that the chaotic flow that is encountered for the non-continuously-modulated stirring protocol is particularly interesting for the thermal mixing of shear-thickening fluids. Generally, non-continuously-modulated stirring protocols have a better efficiency than the two other stirring protocols.

\begin{figure}[htbp]
 \centering
 \begin{center}
  \includegraphics[width=0.49\textwidth]{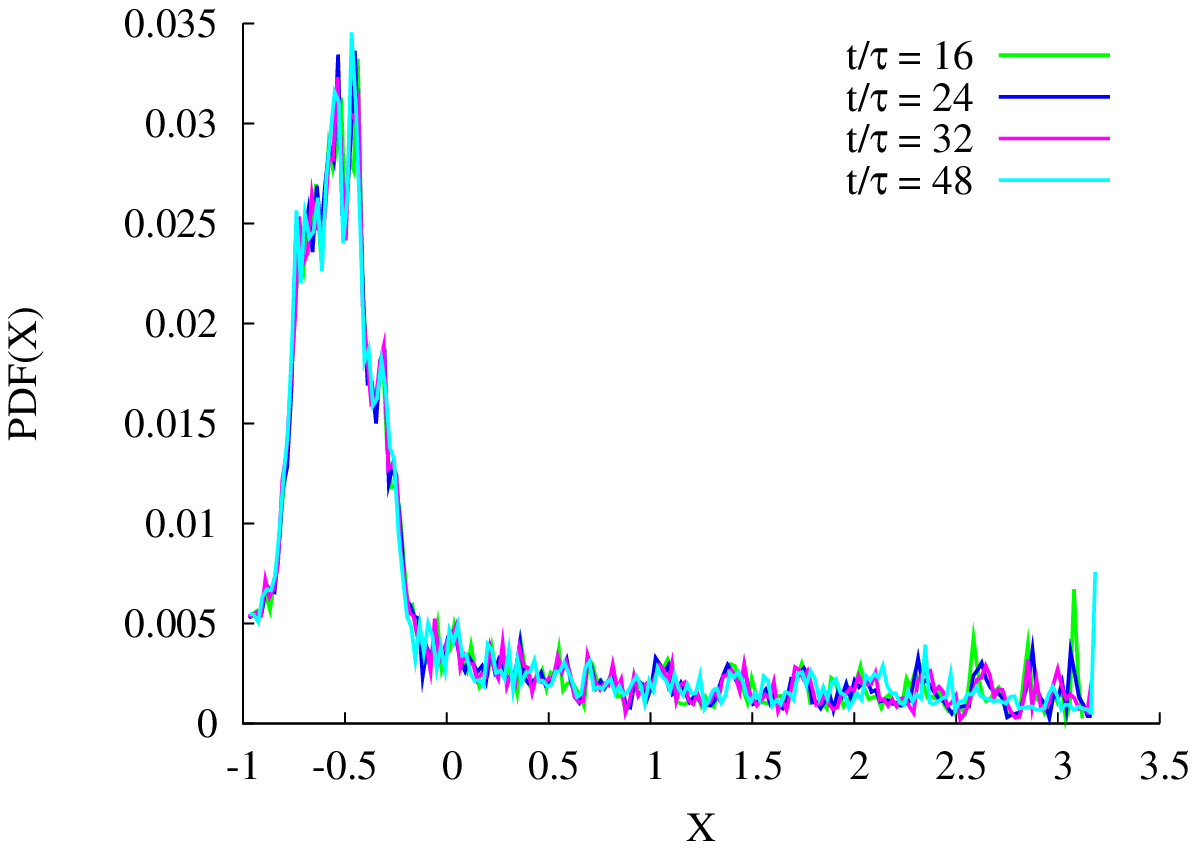}
\includegraphics[width=0.49\textwidth]{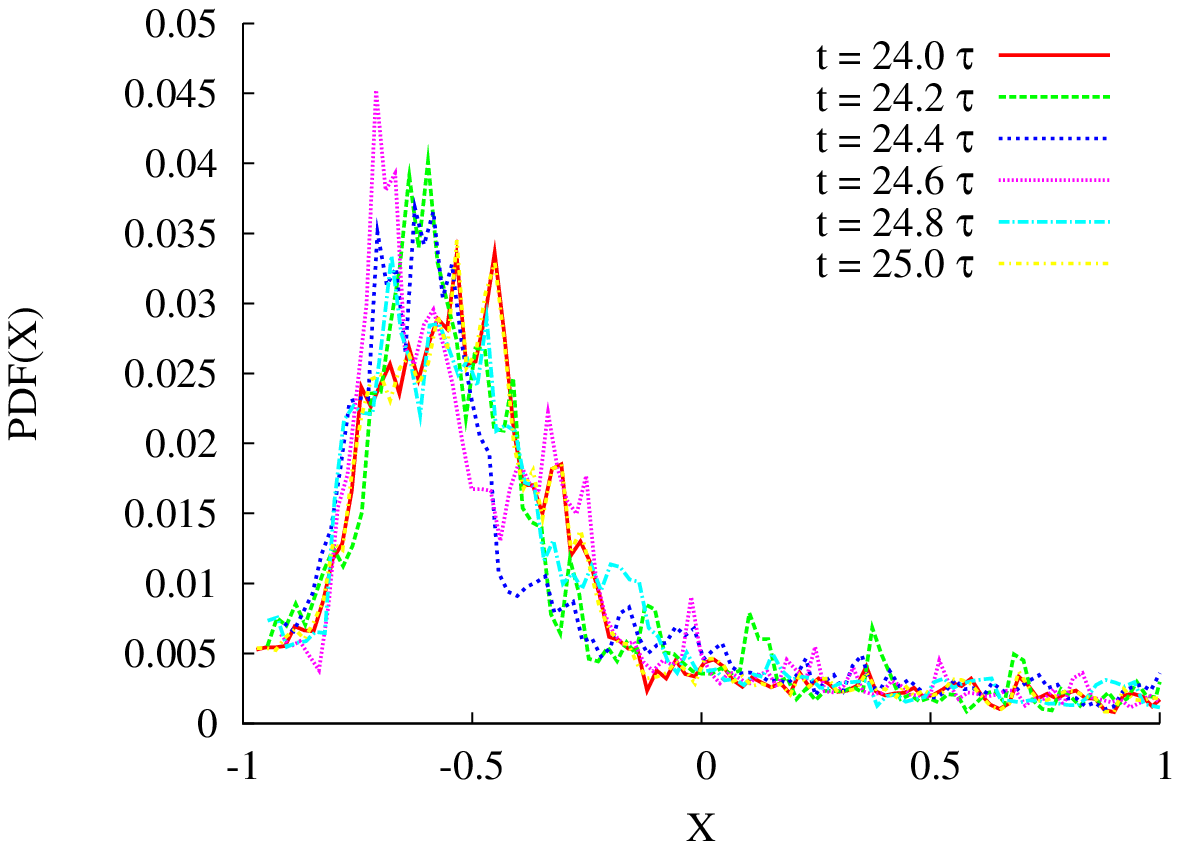}
\\(\textbf{a}) \hspace*{0.45\textwidth}(\textbf{b})
\end{center}
 \caption{PDFs of the rescaled dimensionless temperature, $X$, for the shear-thickening fluid and the non-continuously-modulated stirring protocol \textbf{(a)} at  different period times ($\tau = 30\ s$) and \textbf{(b)} at different times during one period.}
 \label{fig:PDF_X_STK_ALT}
\end{figure}

An important feature of chaotic mixing flows is the periodic resurgence of the same patterns in the scalar field. These patterns appear repeatedly for periodic velocity fields and repeat themselves every period with an exponential decay of the scalar contrast. We showed this phenomenon  for thermal chaotic mixing of a Newtonian fluid in \cite{elomari2009b}. The same characteristic is present for the thermal mixing of shear-thinning and shear-thickening fluids. This feature can be demonstrated by the  PDFs of the rescaled dimensionless temperature:
\begin{equation}
 X = \dfrac{T^*-T^*_m}{\sigma}.
 \label{eq:resdimtemp}
\end{equation}
As an example we show in Fig. \ref{fig:PDF_X_STK_ALT} the PDFs of this quantity in the case of the alternated mixing for the shear-thickening fluid.  These PDFs are superimposed when they are plotted for different times during the mixing process but at the same phase of the period (see Fig. \ref{fig:PDF_X_STK_ALT} (a)). This is the signature of a strange eigenmode \cite{lester2009,pierrehumbert1994,liu2004}, which is characterized by the production of persistent patterns in the flow. These patterns arise from a combination of stretching, folding and thermal diffusion.  The pattern is the same at each periodic time but the amplitude of the dimensionless temperature tends towards 1. In Fig. \ref{fig:PDF_X_STK_ALT} (b), the PDFs are plotted for different times during the same period. From these PDF evolution, we can see that the temperature distributions evolve within the fluid during a period but that it regains its previous form from one period earlier.

\section{Conclusion}
\label{conclusion}
Numerical simulations of the coupled mixing and heating performances that are induced by chaotic advection in a 2D two-rod mixer for non-Newtonian power-law fluids were performed in this study. Three different stirring protocols were chosen: non-modulated, continuous and alternating (non-continuous). The last two were able to give chaotic flow trajectories. In order to study the thermal mixing enhancement mechanism within the fluids, different mixing and energy indicators and statistical tools were used.
According to the wall boundary condition that was considered (constant wall temperature), the following main conclusions can be made based on the obtained results:
\begin{itemize}
\item[$\bullet$] for the non-chaotic and partially chaotic flows, i.e., non-modulated and continuously-modulated stirring, the thermal mixing is more effective for the shear-thinning fluid,
\item[$\bullet$] for all of the rheological fluid behaviors, a high degree of thermal mixing is obtained for the alternating-stirring protocol (fully-chaotic flow),
\item[$\bullet$] for the alternating-stirring protocol, the most effective thermal mixing is obtained for the shear-thickening fluid,
\item[$\bullet$] for the alternating-stirring protocol, cold, poorly-mixed zones exist when a shear-thinning fluid is considered.
\end{itemize}

Further studies will explore the influence of the fluid's yield stress and also the effect of the thermodependence of the complex fluids on the efficiency of thermal chaotic mixing. An additional mid-term objective is to be able to model the flow behavior of the concentrated emulsions \cite{caubet2009,fournanty2008}.


\bibliographystyle{elsarticle-num}


\section*{Appendix A: mesh size dependence study}

In order to choose the appropriate mesh size, giving the best compromise between accuracy and calculation cost, we perform a study of the dependence of the results to the mesh size. The results are presented below for the case of the shear-thickening fluid. We compare the results given for three meshes ($m_1$,  $m_2$ and $m_3$) of increasing sizes : 5960, 10680 and 19380 cells. Both local results and their mean values are compared.

First, we consider two times $t=5\,\tau= 150s$ and $t=10\,\tau= 300s$. We compare the local values of $T^*$ along a vertical cutline passing through the center of the mixer and containing 100 points.  In Tab. \ref{tab:local_err} we report the maximum and the mean value of the relative errors calculated at each point $j$ as $e_j^i = (T^{*i}_j - T^{*3}_j)/T^{*3}_j \times 100$ where $i=1, 2$ stands for meshes $m_1$ and $m_2$ whose results are compared to those of the finest mesh $m_3$. Thus the mean value of the error is $ \bar e^i = \sum_j e_j^i/100$ and the maximum error is $e^i_{max}=\max_j e_j^i$.

\begin{table}[th]
 \begin{center} 
\begin{tabular}{c c c c c}\hline
   mesh            &  $ \bar e(5\tau)$ & $e_{max}(5\tau)$&$ \bar e(10\tau)$&  $e_{max}(10\tau)$\\
               \hline
 $m_1$ &  2.67\%    & 9.60\%    &1.50\% & 4.95\%\\
$m_2$ &   1.44\%    & 5.72\% & 0.94\% & 3.31\%\\
\hline
 \end{tabular}  
\caption{Relative error of the local values of $T^*$ along a vertical cutline at $t=5 \tau$ and $t=10 \tau$. Comparison with the results given by the finest mesh $m_3$. \label{tab:local_err}}
\end{center}
\end{table}
 
 Mean values  $T_m^*$ and $\sigma$ are also compared.  In Tab. \ref{tab:mean_err} we give the maximum and the mean value over time of the relative error to the finest mesh results computed at each instant during the mixing process from $t=2s$ to $t=1500s$.
 
 \begin{table}[th]
 \begin{center} 
\begin{tabular}{c c c c c}\hline
   mesh            &  $ \bar e(\sigma)$ & $e_{max}(\sigma)$&$ \bar e(T_m^*)$&  $e_{max}(T_m^*)$\\
               \hline
 $m_1$ &  2.24\%    & 4.70\%    &0.12\% & 3.52\%\\
$m_2$ &   1.25\%    &2.74\% & 0.05\% & 0.83\%\\
\hline
 \end{tabular}  
\caption{Relative error of mean values of $T^*_m$  and $\sigma$ for $t: 2s \longmapsto 1500s $. \label{tab:mean_err}}
\end{center}
\end{table}

As it can be seen in these two tables, the results given by the mesh $m_2$ are sufficiently close to those given by the  mesh $m_3$ which is two times finner. Hence the mesh $m_2$ was adopted to perform the computations presented in this work.
 
\end{document}